\documentstyle[12pt]{article}
\title{Is quantum field theory a genuine quantum theory? \\
Foundational insights on particles and strings}
\author{Hrvoje Nikoli\'c \\
Theoretical Physics Division, Rudjer Bo\v{s}kovi\'{c} Institute, \\
P.O.B. 180, HR-10002 Zagreb, Croatia \\
{\normalsize hrvoje@thphys.irb.hr} \\
\makebox[1in]{} \\
}
\date{\today}
\begin{document}
\maketitle
\begin{abstract}
Practically measurable quantities resulting from quantum field theory 
are not described by hermitian operators, contradicting one of the
cornerstone axioms of orthodox quantum theory.
This could be a sign that some of the axioms of orthodox
quantum theory should be reformulated.
A non-orthodox reformulation of quantum theory based
on integral curves of particle currents is advocated and possible 
measurable manifestations are discussed. 
The consistency with particle creation and destruction requires 
particles to be extended objects, which can be viewed as a new
motivation for introducing string theory.
Within this reformulation, an indirect low-energy test
of string theory is also possible.
\end{abstract}
\vspace*{0.5cm}

\section{Introduction}

When quantum field theory (QFT) is compared to the so-called 
first quantization of nonrelativistic point particles, it is 
frequently claimed that QFT is nothing but the standard rules of first
quantization applied to different classical degrees of freedom
-- fields. Yet, such a claim is not completely true.
For example, fermionic fields do not really have a classical
counterpart and do not represent quantum observables.
Hermitian operators representing quantum observables can be 
constructed by taking products of fields 
in which the number of fermionic fields is even. 
An example is the energy-momentum tensor $\hat{T}_{\mu\nu}(x)$.
Nevertheless, although such operators correspond to quantities that 
should be measurable in principle, 
they are not quantities measured in practice.
Instead, practically measurable predictions resulting from QFT
are properties of {\em particles}. 
On the other hand, the {\em theoretical} concept of particles in interacting
QFT is rather artificial (see, e.g., \cite{nikmyth} and references 
therein). 

One measurable property of 
particles is their number, which can be described by the
particle number operator $\hat{N}$ constructed from fields.
%
%
%
Still, the number of particles
is not the only property of particles that we measure. Instead, 
we also measure some additional properties of particular particles, such as
their momenta and positions. The momentum operator is not a problem;
it can be defined in QFT, which allows also to reproduce the predictions
of single-particle QM in the momentum space. The problem is 
the {\em position} operator. QFT does not contain a 
position operator. Single-particle relativistic QM does not 
contain it either (see, e.g., \cite{nikmyth} and references therein).
On the other hand, particle detectors in high-energy physics
measure particle positions. In most practical situations this is
not really a problem, because particle-physics phenomenologists have developed
successfull intuitive rules that allow to predict the distribution
of particle positions from scattering amplitudes in the {\em momentum} space.
Still, there is a problem as a matter of principle: If the position
operator is not really defined in QFT, while the position of particles
is a quantity that we really measure when compare the predictions of QFT
with experiments, can we really say that QFT is a genuine quantum theory?

One possible response to this problem is that the position is 
just a classical label, similar to time in nonrelativistic QM.
Such an interpretation is indeed a natural one if one views field
operators $\hat{\phi}({\bf x},t)\equiv\hat{\phi}(x)$ as the fundamental objects
from which everything else should be derived. However, from such a
theoretically natural way of viewing QFT, there is no way to derive
the experimentally confirmed QM rule that the nonrelativistic wave function
$\psi({\bf x},t)$ defines quantum probabilities $|\psi({\bf x},t)|^2$
of particle positions. Are nonrelativistic QM and QFT two
independent theories describing different objects? If so, then why,
in many situations, both theories describe the same particles?
If, on the other hand, nonrelativistic QM is to be derived from
more fundamental QFT, then how to derive the probabilistic 
interpretation of nonrelativistic $\psi({\bf x},t)$ by starting 
from the axioms of QFT? It seems that the standard orthodox
formulation of quantum theory cannot answer these questions.
Therefore, the purpose of this paper is to reformulate some basic
principles of quantum theory, with the intention to obtain a 
self-consistent theory the axioms of which are compatible with all 
existing observations.
As a byproduct, we also find that such a reformulated quantum theory
leads to some testable predictions that cannot be obtained with the
conventional orthodox formulation.

\section{The roles of particles and fields}

Our starting point is the fact that all measurements confirming the 
predictions of QFT are actually measurements of the properties of
{\em particles}. Therefore, particles play the central role in
our reformulation of quantum theory. Still, the success of QFT
cannot be denied, so quantum fields play some role as well, 
but not as fundamental directly measurable quantities, but as
auxiliary mathematical objects from which the properties of 
physical particles are calculated. (From such a point of view,
the expression ``second quantization" seems more suitable than
the expression ``quantum field theory".) 
In the simplest case of spinless uncharged particles, the relation between
quantum fields and particles is encompassed by the equalities of the form
\cite{schw,nikfpl1}
\begin{equation}\label{wf}
\psi(x_1,\ldots,x_n)= S_{\{x_a\}} 
\langle 0| \hat{\phi}(x_1)\cdots\hat{\phi}(x_n) |\Phi\rangle ,
\end{equation}
where $\psi$ is the (unnormalized) $n$-particle wave function, 
$|\Phi\rangle$ is the 
QFT state in the Heisenberg picture (in general, $|\Phi\rangle$ is a 
superposition of states with different numbers of particles $n$),
$|0\rangle$ is the vacuum, 
and $S_{\{x_a\}}$ denotes the symmetrization over all $x_a$,
$a=1,\ldots,n$, needed because the field operators do not commute
for nonequal times. The expression (\ref{wf}) can be easily generalized 
to other types of particles as well, including the fermionic ones
\cite{nikfpl2}. The expression (\ref{wf}) is valid even in the case
of interacting fields \cite{nikfpl1,nikfpl2}, but, for simplicity,
we put emphasis on the free case.

When the field satisfies the Klein-Gordon equation
$
(\partial^{\mu}\partial_{\mu}+m^2)\hat{\phi}(x)=0 
$,
then $\psi$ satisfies $n$ Klein-Gordon equations
\begin{equation}\label{KGn}
(\partial^{\mu}_a\partial_{a\mu}+m^2)\psi(x_1,\ldots,x_n)=0 ,
\end{equation}
one for each $a$. Consequently, the quantity $|\psi|^2$ cannot be
interpreted as a probability density, because such probability would not
be conserved. Instead, one can construct $n$ conserved Klein-Gordon currents
\begin{equation}\label{KGc}
j_{a\mu}=i\psi^*  
\!\stackrel{\leftrightarrow}{\partial}_{\!a\mu}\! \psi , \;\;\;\;
\partial_{a\mu} j^{\mu}_a=0 ,
\end{equation}
but the time components $j_{a0}$ cannot be interpreted as
probability densities either, as they can be negative. 
(Note also that such currents are related to a hermitian uncharged field,
so an interpretation in terms of charge currents would also be 
inappropriate.)
Still, in the nonrelativistic limit, one should recover the standard rule that 
$|\psi|^2$ represents the probability density. This suggests that
{\em the probabilistic interpretation of $\psi$ should be
emergent, rather than fundamental}. This motivates us
to search for a fundamentally 
{\em deterministic} interpretation of $\psi$, from which, 
in the nonrelativistic limit, 
the usual probabilistic interpretation can be derived as an {\em effective} 
theory. 

\section{Bohmian interpretation}

Fortunately, in the nonrelativistic limit, the simplest way to achieve this
is known for more than 50 years: the Bohmian interpretation of
nonrelativistic QM \cite{bohm,bohmPR1,holbook}
provides a fundamentally deterministic explanation of all 
probabilistic predictions of orthodox QM, as long as the 
predictions refer to measurements of observables described by
hermitian operators. Nevertheless, the Bohmian interpretation
is technically more complicated than the orthodox fundamentally 
probabilistic interpretation, which is the main reason that it is
usually ignored by physicists, despite the fact that it may
have practical applications as well \cite{lopreore}.
In this paper, however, the motivation for introduction of 
the Bohmian interpretation is significantly extended. Our goal is not only
to reproduce the predictions of nonrelativistic QM, but also to 
extend the theory to the relativistic regime. To repeat, in this regime
the axioms of orthodox QM are not appropriate, 
as there is no hermitian operator corresponding to the measured
particle positions.

For the case in which the particle creation and destruction is neglected,
the appropriate relativistic generalization of the Bohmian interpretation
is suggested in \cite{durr96} and further developed in \cite{nikfpl3}.
The crucial role is played by the integral curves (see also \cite{nikfol})
of the vector field (\ref{KGc}). The integral curves 
can be represented by functions $X^{\mu}_a(\tau)$ satisfying
\begin{equation}\label{Bohm}
\frac{dX^{\mu}_a(\tau)}{d\tau}=j^{\mu}_a ,
\end{equation}
where $\tau$ is an auxiliary affine parameter along the integral curves.
The functions $X^{\mu}_a(\tau)$ can be viewed either as one curve in the
$4n$-dimensional configuration space, or $n$ synchronized curves
in $4$-dimensional spacetime. This makes Bohmian nonlocality
consistent with the principle of relativistic covariance, so that 
a ``preferred" synchronization between different particles
is chosen dynamically, essentially by the choice of initial conditions
(see also \cite{niklosinj}).
As the currents (\ref{KGc}) may be spacelike at some regions of spacetime,
superluminal particle velocities and motions backwards in time are 
also possible.
Nevertheless, when particle velocities or positions are measured 
appropriately, measured particle velocities can only correspond 
to non-superluminal eigenvalues of the velocity operator, while
multiple positions of a single particle at a single time cannot
be observed \cite{nikfpl1,nikfpl3}. Still, as we discuss below, some
nontrivial measurable manifestations of superluminal velocities 
and motions backwards in time are possible.

\section{Measurable predictions}

As demonstrated in \cite{nikfpl3}, in the relativistic case, the
measurable probabilistic predictions emerging from (\ref{Bohm})         
may differ from more conventional interpretations that do not
involve a calculation of integral curves of the particle           
current. To discuss it, we consider the case of a single
particle with the current $j^{\mu}$.
In general, since we deal with a fundamentally deterministic
theory, there is {\it a priori} no simple mathematical rule 
that gives us a probability density of particle positions.
Thus, the fact that $j^0$
cannot be interpreted as a probability density does not represent
a fundamental problem. 
Nevertheless, in the nonrelativistic approximation
in which $j^0 \propto |\psi|^2$, the usual quantum probability
distribution $|\psi|^2$ emerges as a distribution that corresponds to the 
quantum equilibrium \cite{equilval,equildurr}. In general,
if one knows the distribution at some particular initial time,
one can find the distribution at any other time by calculating
the relativistic particle trajectories for all possible 
initial positions of the particles. Thus,
in the fully relativistic case, one finds a nontrivial measurable prediction
when the interaction with the apparatus that measures particle
positions starts at some particular time $t_1$,
provided that $j_0<0$ at some parts of the $t_1$-hypersurface.
Let us briefly explain it. (For more details we refer the reader to \cite{nikfpl3}.)
Negative $j_0$ corresponds to a particle moving backwards in time.
However, since the wave function (\ref{wf}) contains only positive frequencies,
the motion backwards in time is only a transient phenomenon for a given 
particle trajectory. Thus, in order for a particle to move backwards in time
at $t_1$, it must come to $t_1$ from larger times, by changing 
the direction of motion from a motion forward in time to a motion backwards in time.
Clearly, this change of the direction of motion must happen at some time
$t>t_1$. However, since the 4-velocity on a given spacetime point
is unique, such a motion must also involve a motion is space, not only in time.
On the other hand, a typical measuring apparatus that measures a particle
position does not allow a motion in space \cite{bohm,bohmPR1,holbook}.
Consequently, since the measuring apparatus is turned on for $t>t_1$,
the effect of measurement is that the particle cannot arrive at points
on the $t_1$-hypersurface at which $j_0$ is negative.
Thus, one finds the probability density \cite{nikfpl3}
\begin{equation}\label{pred}
\rho({\bf x},t_1)=\left\{
\begin{array}{ll}
j_0({\bf x},t_1) & \mbox{on $\Sigma'$} , \\
0 & \mbox{on $\Sigma^+\cup\Sigma^-$} ,
\end{array}
\right.
\end{equation}
where $\Sigma^-$ is the part of the $t_1$-hypersurface at which
$j_0<0$, $\Sigma^+$ is the part of the $t_1$-hypersurface
that is connected to $\Sigma^-$ by Bohmian trajectories
for which $t<t_1$, and $\Sigma'$ represents all other points of the
$t_1$-hypersurface. 
We emphasize that
this measurable result cannot be obtained without
calculating the trajectories. The most surprising feature of this prediction
is that the particle cannot be found at some positions at which
the wave function does not vanish. 
This could be observed experimentally 
for a state described by a coherent superposition
of two significantly different frequencies $\omega_1$ and $\omega_2$.
Unfortunately, such an experiment requires very fast switching
(lasting much shorter than $|\omega_1-\omega_2|^{-1}$) 
of the measuring apparatus at $t_1$, which turns out to be difficult
to achieve with the present-day technology, given that 
the frequencies are sufficiently different to allow for a measurable effect
\cite{genov}. Nevertheless, different (hopefully more feasible)
measurable manifestations based on
explicit calculations of particle trajectories are also conceivable \cite{nikolprob}.

\section{Particle creation and destruction and the role of strings}

One of the appealing features of the Bohmian interpretation 
of quantum particles is the fact that
it avoids discontinuous jumps corresponding to the so-called wave-function
collapses. All evolution is fully continuous, described by the
deterministic wave equation such as (\ref{KGn})
and the attributed deterministic particle trajectories described by
(\ref{Bohm}). However, the problem is to reconcile it with the
fact that particles may also get created or destructed. 
One possibility is to introduce singular points at which particle
trajectories begin or end \cite{durrprl}. Another possibility is to
introduce an additional continuously changing property of particles
called effectivity \cite{nikfpl1}. Still, both possibilities seem rather
artificial. In addition, both possibilities require an explicit use of
QFT, which seems unappealing for a theory that is fundamentally 
supposed to be a theory of particles, rather than fields. So, how
to describe particle creation and destruction in a continuous way
without postulating new quantities such as effectivities?
Below we argue that the requirement of continuous creation and destruction
of particles naturally leads to {\em string theory}.
In this way, string theory can be viewed as a natural consequence
(or even a prediction) of the Bohmian interpretation.

Consider a decay of one particle with the trajectory $X_1^{\mu}(\tau)$ into 
two new particles with the trajectories $X_2^{\mu}(\tau)$ and
$X_3^{\mu}(\tau)$.
All together, we have 3 trajectories $X_a^{\mu}(\tau)$, $a=1,2,3$.
How to describe the transition from $X_1^{\mu}(\tau)$ to 
$X_2^{\mu}(\tau)$ and $X_3^{\mu}(\tau)$ in a continuous way without singular
points at which these trajectories begin or end? The most obvious
answer is by allowing the label $a$ to attain not only discrete 
integer values, but also real values that can continuously interpolate
between the integer ones.
In general, there may be $p$ independent real variables
$\sigma_1,\ldots, \sigma_p$, so the trajectories 
$X_a^{\mu}(\tau)$ get replaced by more general functions 
$X^{\mu}(\sigma_1,\ldots, \sigma_p,\tau)$. Geometrically, such general
functions represent $p$-dimensional extended objects evolving with 
``time" $\tau$.
Further, if such objects are considered to be fundamental, one requires
that their quantization does not lead to unsurmountable divergences,
which excludes the cases $p>1$ \cite{GSW}. 
This leads to
functions $X^{\mu}(\sigma,\tau)$, which are nothing but
strings in spacetime. Indeed, strings are already known to be a good 
candidate for a more fundamental theory that should replace the 
usual theory of particles and fields \cite{GSW,polc}. 
In fact, it has already been argued that, if strings are assumed to be
the fundamental objects, the Bohmian interpretation 
appears as a natural interpretation  
\cite{nikstr1,nikstr2,nikstr3}. Our result above shows also the opposite:
{\em if the Bohmian interpretation is assumed, 
then strings appear as natural objects}. Loosely speaking,
strings are a prediction of the Bohmian interpretation. 

In string theory, one can describe string creation and destruction without
introducing an analog of fields. (String field theory is also a 
possible modification of string theory, but 
such a modification does not seem to be 
necessary and there are indications that it may not be the correct way
to go \cite{polc2}.)
The $n$-particle wave function
$\psi(x_1,\ldots,x_n)$ generalizes to the bosonic string wave functional
$\Psi[X(\sigma)]\equiv\Psi[X]$. 
Allowing functions $X^{\mu}(\sigma)$ to be discontinuous, 
the wave functional $\Psi[X]$ describes not only 1-string
states, but also $n$-string states, as well as amplitudes for 
processes in which the number of strings changes by string splitting 
\cite{nikstr3}. There is no need for a stringy analog of (\ref{wf}).
In the case of superstrings, the wave functional 
further generalizes to $\Psi[X,M]$, where $M(\sigma)$ are indices
in the Hilbert space attributed to the operator corresponding to the 
anticommuting superstring coordinate $\psi^{\mu}(\sigma)$ \cite{nikstr3}.  
The particle current (\ref{KGc}) generalizes to the
superstring current \cite{nikstr3}
\begin{equation}\label{strc}
J_{\mu}[X;\sigma)=i\int [dM] \,  \Psi^*[X,M]
\frac{
\!\stackrel{\leftrightarrow}{\delta}\! }
{\delta X^{\mu}(\sigma)}
\Psi[X,M] .
\end{equation}
Eq.~(\ref{Bohm}) generalizes to 
\begin{equation}\label{Bohms}
\frac{\partial X^{\mu}(\sigma,\tau)}{\partial\tau}=J^{\mu}[X;\sigma) , 
\end{equation}
which describes the continuous deterministic motion of strings, including the 
string splitting \cite{nikstr3}.

\section{Low-energy test of string theory}

To discuss possible measurable manifestations of our non-orthodox 
reformulation of string theory, it is convenient to study the 
implications on particles of the Standard Model. The main idea 
is to consider an appropriate low-energy string state $\Psi$
and its effective description in terms of the corresponding
particle wave function $\psi$. Then from (\ref{strc}) we obtain
a particle current generalizing (\ref{KGc}) to higher spins, which
leads to non-orthodox measurable predictions such as (\ref{pred}). 
For example, the photon wave function is obtained from a
massless vector state of the open string, leading to a 
general 1-photon wave function
\begin{equation}\label{ph}
\psi(x)=\sum_{k}\sum_{s} c_s(k) e^{-ik\cdot x} ,
\end{equation}
where $k^{\mu}$ is the particle momentum, 
$s$ labels $D-2$ independent polarization vectors $\zeta^{\alpha}_s(k)$
satisfying $k_{\alpha}\zeta^{\alpha}_s(k)=0$,
and $c_s(k)$ are arbitrary coefficients
corresponding to an arbitrary superposition of states
with various momenta and polarizations. The corresponding
low-energy 1-photon current resulting from (\ref{strc})
takes the same form as (\ref{KGc}), that is
\begin{equation}\label{KGc1}
j_{\mu}=i\psi^*
\!\stackrel{\leftrightarrow}{\partial}_{\!\mu}\! \psi . 
\end{equation}
This should be contrasted with the
1-photon current that would naturally emerge directly from QFT 
\cite{nikfol}
\begin{equation}\label{KGcf}
j_{\mu}^{\rm field} =i\psi_{\alpha}^*
\!\stackrel{\leftrightarrow}{\partial}_{\!\mu}\! \psi^{\alpha} , 
\end{equation} 
where $\psi^{\alpha}$ is obtained from a spin-1 analog of (\ref{wf}), 
leading to
\begin{equation}\label{phf}
\psi^{\alpha}(x)=\sum_{k}\sum_s c_s(k) 
\zeta^{\alpha}_s(k) e^{-ik\cdot x} .
\end{equation}
In general, the predictions from these two different 1-photon currents
differ, provided  that
the superposition contains contributions from various non-collinear 
momenta $k$. In principle,
this difference could be used for an indirect low-energy test
of string theory.

A more dramatic difference between the string-based particle
current and the field-based particle current occurs for spin-$\frac{1}{2}$
particles. The field-based 1-electron current is the Dirac current
\begin{equation}\label{fercurf}
j_{\mu}^{\rm field} =\psi^{\dagger}\gamma^0\gamma_{\mu}\psi ,
\end{equation}
where $\psi$ is the spinor with the c-number valued components $\psi_m$,
while $\gamma_{\mu}$ are the Dirac matrices. The string-based
low-energy 1-electron current resulting from (\ref{strc}) is
\begin{equation}\label{fercurs}
j_{\mu} =i\psi^{\dagger}
\!\stackrel{\leftrightarrow}{\partial}_{\!\mu}\! \psi ,
\end{equation}
where the index $m$ originates from $M$ \cite{nikstr3}.
In the ``field-frame" the current (\ref{fercurs}) does not seem
to transform as a vector, but in the ``string-frame" it is the
Dirac current (\ref{fercurf}) that does not seem to transform
as a vector. In \cite{nikstr3} it is interpreted as an 
existence of a preferred foliation of spacetime at the level
of effective field theory. The most notable difference between 
these two currents is the fact that $j_0^{\rm field}$ 
in (\ref{fercurf}) cannot be
negative, so nontrivial probability distributions such as
(\ref{pred}) cannot emerge from (\ref{fercurf}). 
Therefore, the experimental search for such nontrivial probability 
distributions of electrons would be an indirect low-energy test of
string theory. 

A more specific proposal for testing the difference between 
probabilistic predictions of (\ref{fercurf}) and (\ref{fercurs})
is discussed in detail in \cite{niktest}. Let us briefly review it here.
In 4 dimensions,
the wave function of an ultrarelativistic electron with spin $+\frac{1}{2}$
having energy $\omega_p$ and moving in the $x$-direction is
\begin{equation}\label{spinor2}
 \varphi_p=\frac{1}{\sqrt{2}}
\left( 
\begin{array}{c}
 1 \\ 0 \\ 0 \\ 1
\end{array}
\right) e^{-i\omega_p(t-x)} .
\end{equation}
The main idea is to prepare the electrons in a superposition
of two different equally probable frequencies $\omega_1$ and $\omega_2$.
Thus, the wave function in  (\ref{fercurf}) is
\begin{equation}\label{supD}
\psi=\frac{1}{\sqrt{2}}(\varphi_1+\varphi_2) ,
\end{equation}
where $\varphi_1$, $\varphi_2$ denote spinors (\ref{spinor2})
with frequencies $\omega_1$, $\omega_2$, respectively.
On the other hand, the Klein-Gordon-like current (\ref{fercurs})
requires a Klein-Gordon normalization of the wave function,
so instead of (\ref{supD}), the corresponding wave function in 
(\ref{fercurs}) is 
\begin{equation}\label{supKG}
\psi=\frac{1}{\sqrt{2}} \left( \frac{\varphi_1}{\sqrt{2\omega_1}}
+\frac{\varphi_2}{\sqrt{2\omega_2}} \right) .
\end{equation}
Now we put the detector of electrons at a fixed position, say $x=0$.
The probability density resulting from (\ref{fercurf}) with (\ref{supD}) is
\begin{equation}\label{rhoDm}
j_0^{\rm field}(t)=1+ {\rm cos}[(1-\eta)\omega_1 t]  ,
\end{equation}
where $\eta\equiv \omega_2/\omega_1$.
Similarly, the probability density resulting from (\ref{fercurs}) with (\ref{supKG}) is
\begin{equation}\label{rhoKGm}
|j_0(t)|=\left| 1+ 
\frac{1+\eta}{2\sqrt{\eta}} \, {\rm cos}[(1-\eta)\omega_1 t] \right| .
\end{equation}
For $\omega_1 \neq \omega_2$, (\ref{rhoDm}) and (\ref{rhoKGm})
are different. The source of the difference lies in the fact
that $j_0^{\rm field}$ defined by (\ref{fercurf}) does not involve
a time-derivative while $j_0$ defined by (\ref{fercurs}) does,
which causes substantial differences for superpositions of
different frequencies.
As discussed in more detail in \cite{niktest}, the oscillatory patterns
(\ref{rhoDm}) and (\ref{rhoKGm}) can, at least in principle, be distinguished
experimentally.

\section{Conclusion}

To conclude, our results suggest that quantum field theory is neither
a quantum theory in the orthodox sense, nor a field theory in a 
naive sense.
It is not an orthodox quantum theory because practically measurable 
quantities resulting from QFT are particle positions, which are
not described by hermitian operators. It is not a field theory
in a naive sense
because the fundamental measurable quantities at low energies are
particles rather than fields. 
The fact that a relativistic position operator does
not exist lead us to argue that the usual quantum probabilities are emergent,
while the fundamental dynamics is deterministic.
This suggests a relativistic Bohmian-like formulation
of quantum theory, which, in the relativistic regime, 
leads to testable predictions. 
To make the existence of Bohmian particle trajectories consistent
with particle creation and destruction without leading to
singular points at which trajectories begin or end, it is natural
to generalize pointlike particles to extended objects. This leads
to a new derivation of string theory based on the assumption
of Bohmian mechanics, which allows testable low-energy predictions
of string theory as well.

\section*{Acknowledgements}
This work was supported by the Ministry of Science of the
Republic of Croatia under Contract No.~098-0982930-2864.

\end{document}